\input epsf.tex
\documentclass[12pt]{article}
\usepackage{graphicx}
\usepackage{psfig,picinpar,floatflt,amssymb,epsf}
\csname @addtoreset\endcsname{equation}{section}

\def\bseq{\begin{subequation}}  
\def\eseq{\end{subequation}}
\def\bsea{\begin{subeqnarray}}  
\def\esea{\end{subeqnarray}}

\newcommand{\beq}{\begin{equation}}
\newcommand{\eeq}{\end{equation}}
\newcommand{\bea}{\begin{eqnarray}}
\newcommand{\eea}{\end{eqnarray}}
\newcommand{\ena}{\end{eqnarray}}

\renewcommand{\a}{\alpha}
\renewcommand{\b}{\beta}

\newcommand{\pa}{\partial}

\newcommand{\phib}{\bar{\phi}}
\newcommand{\Phib}{\bar{\Phi}}
\newcommand{\adot}{\dot{\alpha}}
\newcommand{\bdot}{\dot{\beta}}

\newcommand{\thb}{\bar{\theta}}

\def\Mb{\kern 2pt\mathchoice
        {
         \vbox{\hrule width10pt height 0.4pt depth 0pt
         \kern 1.2pt\hbox{\kern -2pt$\displaystyle M$}}}
        {
         \vbox{\hrule width10pt height 0.4pt depth 0pt
         \kern 1.2pt\hbox{\kern -2pt$\textstyle M$}}}
        {
\vbox{\hrule width6pt height 0.4pt depth 0pt
         \kern 1.0pt\hbox{\kern -2pt$\scriptstyle M$}}}
        {
         \vbox{\hrule width5pt height 0.4pt depth 0pt
         \kern 0.8pt\hbox{\kern -2pt$\scriptscriptstyle M$}}}}

\def\Sb{\kern 2pt\mathchoice
        {
         \vbox{\hrule width6pt height 0.4pt depth 0pt
         \kern 1.2pt\hbox{\kern -2pt$\displaystyle S$}}}
        {
         \vbox{\hrule width6pt height 0.4pt depth 0pt
         \kern 1.2pt\hbox{\kern -2pt$\textstyle S$}}}
        {
         \vbox{\hrule width3.5pt height 0.4pt depth 0pt
         \kern 1.0pt\hbox{\kern -2pt$\scriptstyle S$}}}
        {
         \vbox{\hrule width3pt height 0.4pt depth 0pt
         \kern 0.8pt\hbox{\kern -2pt$\scriptscriptstyle S$}}}}

\def\Rb{\kern 2pt\mathchoice
        {
         \vbox{\hrule width5.5pt height 0.4pt depth 0pt
         \kern 1.2pt\hbox{\kern -2.5pt$\displaystyle R$}}}
        {
         \vbox{\hrule width5.5pt height 0.4pt depth 0pt
         \kern 1.2pt\hbox{\kern -2.5pt$\textstyle R$}}}
        {
         \vbox{\hrule width3.5pt height 0.4pt depth 0pt
         \kern 1.0pt\hbox{\kern -2.2pt$\scriptstyle R$}}}
        {
         \vbox{\hrule width3pt height 0.4pt depth 0pt
         \kern 0.8pt\hbox{\kern -2.2pt$\scriptscriptstyle R$}}}}

  \def\pp{{\mathchoice
      {
          \kern 1pt%
          \raise 1pt
          \vbox{\hrule width5pt height0.4pt depth0pt
            \kern -2pt
            \hbox{\kern 2.3pt
              \vrule width0.4pt height6pt depth0pt
              }
            \kern -2pt
            \hrule width5pt height0.4pt depth0pt}%
            \kern 1pt
       }
        {
          \kern 1pt%
          \raise 1pt
          \vbox{\hrule width4.3pt height0.4pt depth0pt
            \kern -1.8pt
            \hbox{\kern 1.95pt
              \vrule width0.4pt height5.4pt depth0pt
              }
            \kern -1.8pt
            \hrule width4.3pt height0.4pt depth0pt}%
            \kern 1pt
        }
        {
          \kern 0.5pt%
          \raise 1pt
          \vbox{\hrule width4.0pt height0.3pt depth0pt
            \kern -1.9pt  
            \hbox{\kern 1.85pt
              \vrule width0.3pt height5.7pt depth0pt
              }
            \kern -1.9pt
            \hrule width4.0pt height0.3pt depth0pt}%
            \kern 0.5pt
        }
        {
          \kern 0.5pt%
          \raise 1pt
          \vbox{\hrule width3.6pt height0.3pt depth0pt
            \kern -1.5pt
            \hbox{\kern 1.65pt
              \vrule width0.3pt height4.5pt depth0pt
              }
            \kern -1.5pt
            \hrule width3.6pt height0.3pt depth0pt}%
            \kern 0.5pt
        }
    }}

  \def\mm{{\mathchoice
               {
                 \kern 1pt
           \raise 1pt    \vbox{\hrule width5pt height0.4pt depth0pt
                  \kern 2pt
                  \hrule width5pt height0.4pt depth0pt}
                 \kern 1pt}
               {
                \kern 1pt
           \raise 1pt \vbox{\hrule width4.3pt height0.4pt depth0pt
                  \kern 1.8pt
                  \hrule width4.3pt height0.4pt depth0pt}
                 \kern 1pt}
               {
                \kern 0.5pt
           \raise 1pt
                \vbox{\hrule width4.0pt height0.3pt depth0pt
                  \kern 1.9pt
                  \hrule width4.0pt height0.3pt depth0pt}
                \kern 1pt}
               {
               \kern 0.5pt
         \raise 1pt  \vbox{\hrule width3.6pt height0.3pt depth0pt
                  \kern 1.5pt
                  \hrule width3.6pt height0.3pt depth0pt}
               \kern 0.5pt}
               }}

\def\pd{{\kern0.5pt
           + \kern-5.05pt \raise5.8pt\hbox{$\textstyle.$}\kern
0.5pt}}

\def\pmd{{\kern0.5pt
          \pm \kern-5.05pt
\raise6.3pt\hbox{$\textstyle.$}\kern1.5pt}}

\def\md{{\mathchoice
   {
      {{\kern 1pt - \kern-6.2pt \raise5pt\hbox{$\textstyle.$}\kern
1pt}}}
    {
      {{\kern 1pt - \kern-6.2pt \raise5pt\hbox{$\textstyle.$}\kern
1pt}}}
    {
      {\kern0.5pt - \kern-5.05pt
\raise3.4pt\hbox{$\textstyle.$}\kern0.5pt}}
    {
      {\kern0.5pt - \kern-5.05pt
\raise3.4pt\hbox{$\textstyle.$}\kern0.5pt}}}}

\begin{document}

\begin{titlepage}

\begin{center}
\vglue .06in
{\Large\bf Nonanticommutative superspace \\
and N= 1/2 WZ model}
\\[.45in]
Marcus T. Grisaru\footnote{grisaru@physics.mcgill.ca}\\
{\it Physics Department, McGill University \\
Montreal, QC Canada H3A 2T8 }
\\
[.2in]
Silvia Penati\footnote{silvia.penati@mib.infn.it} ~and~
Alberto Romagnoni\footnote{alberto.romagnoni@mib.infn.it}\\
{\it Dipartimento di Fisica dell'Universit\`a degli studi di
Milano-Bicocca,\\
and INFN, Sezione di Milano, piazza della Scienza 3, I-20126 Milano,
Italy}\\[.8in]

{\bf ABSTRACT}\\[.0015in]
\end{center}

In these proceedings we review the main results concerning superspace 
geometries with nonanticommutative spinorial variables and field theories
formulated on them. In particular, we report on the quantum properties of
the WZ model formulated in the $N=1/2$ nonanticommutative superspace.

${~~~}$ \newline

\vskip 30pt
\begin{center}
{\em Contribution to the proceedings of the Copenhagen RTN workshop,\\
15--20 September 2003}  
\end{center}

\end{titlepage}

\section{Introduction}

Since the famous paper by Seiberg and Witten \cite{SW} it has been known that
the low--energy dynamics of D3--branes in flat space, in the presence of
a constant Neveu--Schwarz magnetic background is described by a $N=4$ 
supersymmetric Yang--Mills theory defined on a noncommutative 4d spacetime,
$[ x^\mu , x^\nu ] = i \Theta^{\mu\nu}$. 
Previous indications about the possibility to have noncommutative geometry 
from string theory or M(atrix) theory can be found in \cite{Connes}.

Supersymmetry plays a fundamental role in string theory. In particular, in 
the Green--Schwarz \cite{GSW} or Berkovits \cite{Berkovits} formulations of 
superstring  
the target space supersymmetry is made manifest and the superstring 
sigma-model action describes the dynamics of the 
coordinate variables of a ten dimensional superspace.
It is then natural to investigate the possibility of embedding
noncommutativity in superspace and 
defining superspace geometries where the non(anti)commutativity is extended
to the spinorial coordinates. 

Independently of a string context, a number of papers \cite{VAN1, FL, us1}
have studied the possibility to define such supergeometries.
In \cite{us1} a systematic construction of the most general nonanticommutative
(NAC) superspace has been done by studying the compatibility 
between noncommutativity and supersymmetry. Among the results of that paper, it
is important to mention the following: It has been proven that in 
Euclidean $N=(1,1)$ four dimensional superspace a NAC generalization 
which is associative and compatible with supersymmetry can be given where 
nontrivial anticommutators 
$\{ \theta^{\a}_i , \theta^{\b}_j \}$ are turned on.  
Since the $N=1$ euclidean superspace (rigorously called
$N=(\frac12, \frac12)$ superspace) can be defined by 
a suitable truncation of $N=(1,1)$, one
obtains the NAC generalization of euclidean $N=(\frac12, \frac12)$ superspace
with $\{ \theta^{\a} , \theta^{\b} \} = C^{\a \b}$ constant.
As proven in \cite{us1}, the possibility to turn on nontrivial
anticommutation relations for the spinorial coordinates by keeping the 
algebra associative is peculiar of the {\em euclidean} superspace. 
Associativity in Minkowski necessarily requires the spinors 
to be ordinary grassmannian variables.

In a string theory context, the main question is whether 
particular string configurations exist which, in the low energy
limit, give rise to supersymmetric field theories defined on such superspace
geometries. This question has been recently answered in a number
of remarkable papers \cite{OV, Seiberg, BS} where it has been shown that 
in the low energy limit a type IIB string with D3--branes 
compactified on a Calabi--Yau three--fold, in the 
presence of a self-dual, constant 
graviphoton background $F_{\a \b}$ gives rise to
the NAC superspace geometry $\{ \theta^{\a} , \theta^{\b} \}
= {\a'}^2 F_{\a \b}$. We note that the graviphoton is taken to 
be self--dual in order to avoid any back--reaction on the metric which
remains flat. On the other hand, the graviphoton can be taken self--dual
only in {\em euclidean} signature, consistently with the fact that only 
euclidean superspace allows for $\{ \theta^{\a} , \theta^{\b} \} \neq 0$.
Similar results in ten dimensions have been obtained in 
\cite{VAN2}.
   
In \cite{Seiberg} the investigation of field theories defined on such 
geometries has been initiated. This requires the definition of a suitable
graded Poisson structure associated to the nontrivial algebra 
$ \{ \theta^{\a} , \theta^{\b} \}$ and a corresponding 
star product in the space of smooth functions  defined on superspace. 
As we will describe later, there may be different  NAC 
generalizations of the $N=(\frac12, \frac12)$ 
superspace according to which representation 
we choose for the supersymmetry charges and for the corresponding spinorial 
covariant derivatives. Moreover, there is more than one star product 
compatible with $\{\theta^\a, \theta^\b \} \neq 0$.

The important observation is that in some NAC generalizations the
algebra of superfields does not respect the complete $N=1$ supersymmetry.
From \cite{Seiberg} and subsequent discussions \cite{FS, Ivanov} it appears 
that a NAC but susy preserving generalization can be done only at the expense
of a consistent definition of chirality.   
On the other hand, if we insist on keeping the ordinary definition 
of chirality and require the star product of two chirals to be chiral, then
we are forced to dress the algebra of superfields with a star product which
breaks half of the supersymmetry \cite{Seiberg}. We are then led to consider
field theories defined on $N=(\frac12, 0)$ superspace (briefly called 
$N=1/2$ superspace).  

The simplest model one can study on such NAC geometries is the $N=1/2$
generalization of the WZ model describing the dynamics of a scalar 
superfield self-interacting with a cubic superpotential $\Phi \ast \Phi \ast
\Phi$. 
At the classical level it turns out to be 
the ordinary WZ model perturbed by a soft-breaking term $C^2 \int d^4x
F^3(x)$, where $C^2$ is the square of the anticommutation parameter and
$F$ is the highest component of the chiral superfield.
At the quantum level, renormalizability properties of the model have been 
studied in \cite{BFR, Ter, us2}. The model as it stands is not renormalizable 
since at one--loop a divergent contribution arises proportional to 
$C^2 \int d^4x F^2(x)$. However, in \cite{us2} it has been proven that
the addition of this extra term in the classical lagrangian makes the
theory renormalizable up to two loops. 
Further investigations have been carried on 
and a proof of the renormalizability of the model with the extra $F^2$
term at any order of perturbation theory has been eventually given in 
\cite{BFR2}.

We note that the renormalizability of the NAC WZ model is not obvious 
from power
counting due to the appearance of a dimensionful constant (the anticommutation
parameter $C^{\a\b}$) and, consequently, of irrelevant deformations with
scaling dimensions greater than four. However, in \cite{rey} a nice argument 
has been given to prove the renormalizability of the model by power counting.
It is based on the observation that in euclidean space, due
to the lack of  h.c. relations between the spinorial variables $\theta^{\a}$ 
and $\bar{\theta}^{\adot}$ one can make a reassignement of the scale
dimensions of the spinorial coordinates in such a way to have $C^{\a\b}$
dimensionless and the associated $F^3$ deformation as a marginal 
operator.       

In the next Section we review the construction of the $N=1/2$ superspace
and discuss the issue of half--breaking of supersymmetry. In Section 3 
we report the results of \cite{us2} about the two--loop renormalization 
of the NAC WZ model and discuss 
as an extra $F^2$ deformation is needed to guarantee the renormalizability 
of the model. Finally, in the conclusions we mention further developments and
possible future lines of investigation.

\section{The $N=1/2$ nonanticommutative superspace}

Following \cite{us1} one can find the most general algebra for 
the coordinates of a flat superspace compatible with supersymmetry by
imposing the covariance of the fundamental algebra under translations 
and supertranslations.  
If we work in Minkowski signature, the
extra condition for the algebra of the coordinates to be associative brings in
quite severe constraints which allow, as the only nontrivial
commutators, $[x, \theta]$,  $[x, \thb]$ and $[x,x]$. However, it was shown in
\cite{us1} that  euclidean signature is less restrictive and a NAC superspace
with $\{ \theta, \theta \}$ different from zero can be defined consistently
with associativity.

Rigorously, a superspace with euclidean signature can be defined only when
extended susy is present because of the impossibility of assigning consistent
reality conditions for the pair of Weyl fermions $\theta_\a$, $\thb_{\adot}$
(for a detailed review on the subject see for instance \cite{vanproeyen}). 
However, in the $N=1$ case one can still define a superspace with euclidean
signature by temporarly doubling the fermionic degrees of freedom and
choosing nonstandard conjugation rules among them (for a detailed discussion, 
see \cite{Ivanov}). 

We review the results of \cite{us1} on $N=(1,1)$ euclidean superspace 
in view of the fact that eventually we will make a truncation to $N=(\frac12, 
\frac12)$. We describe the $N=(1,1)$ superspace by coordinates
$(x^{\a \adot}, \theta^\a, \thb^\a, \theta^{\adot}, \thb^{\adot})$
subject to the complex conjugation conditions
$(\theta^\a)^{\ast} =  \theta^\b \epsilon_{\b \a}$, 
$(\thb^\a)^{\ast} =  \thb^\b \epsilon_{\b \a}$ 
and the same for dot variables. These are not the standard conjugation
rules for spinorial coordinates since the $\ast$--operation squares to 
$-1$. However, these are the conjugation rules compatible with truncation
to $N=(\frac12, \frac12)$. The important point is that in euclidean
signature there are no h.c. relations between $\theta^\a$ and $\theta^{\adot}$.

The structure of the NAC algebra depends on the representation we choose 
for susy charges and covariant derivatives. In \cite{us1} we chose a 
nonchiral representation which brought to a NAC geometry where
the algebra of covariant derivatives and supercharges were both deformed.
In \cite{Seiberg} an alternative proposal was made which uses 
the chiral representation for derivatives and charges (we consider 
only the left sector and use the conventions of \cite{superspace})
\bea
&& Q_{\adot} = i ( \pa_{\adot} - i \theta^{\a} \pa_{\a \adot} )
\quad , \quad Q_{\a} = i \pa_{\a}
\nonumber \\
&& D_{\adot} =  \pa_{\adot} \qquad \qquad \qquad , \quad
D_{\a} = \pa_{\a} + i \theta^{\adot} \pa_{\a \adot}
\label{DQ}
\eea
In principle the NAC algebra consistent with susy and associativity is
of the form
\bea
&& \{ \theta^\a , \theta^\b \} = 2 C^{\a \b} \qquad \{ \theta^{\adot} ,
\theta^\b \}=
\{ \theta^{\adot} , \theta^{\bdot} \} =0
\nonumber \\
&& [ x^{\a\adot} , \theta^\b ] = -2 i  C^{\a\b} \theta^{\adot}
\qquad [ x^{\a\adot} , x^{\b\bdot} ] = 2 \theta^{\adot} C^{\a\b} \theta^{\bdot}
\label{NC20}
\eea
but a suitable change of variable
$ y^{\a\adot} = x^{\a\adot} - i \theta^\a \theta^{\adot}$
avoids dealing with noncommuting $x$'s. Therefore the superspace coordinates
$(y^{\a \adot}, \theta^\a, \thb^\a, \theta^{\adot}, \thb^{\adot})$
satisfy
\beq
\{ \theta^\a , \theta^\b \} = C^{\a \b} \qquad \qquad {\rm the ~rest} = 0
\label{NAC1}
\eeq
In this case the algebra of the covariant spinor derivatives is not modified,
while
\beq
\{Q_\a , Q_\b\}_\ast = 0 \quad , \quad  \{Q_\a , Q_{\adot} \}_\ast =
i \pa_{\a \adot}\quad , \quad
\{Q_{\adot} , Q_{\bdot} \}_\ast = 2 C^{\a \b} \pa_{\a \adot} \pa_{\b \bdot}
\label{NC30}
\eeq
Therefore the supersymmetry is explicitly broken to $N=(\frac12, 0)$
\cite{Seiberg} on the class of smooth functions defined on this superspace.
We note that the susy-breaking term is quadratic in the bosonic
derivatives, so
it does not spoil the previous statement about consistency of ({\ref{NAC1}) 
with supersymmetry invariance of the fundamental algebra of the coordinates.

Following Seiberg we realize the NAC geometry on the smooth superfunctions
by introducing the nonanticommutative
(but associative) product
\beq
\phi \ast \psi = \phi e^{- \overleftarrow{\pa}_\a C^{\a \b}
\overrightarrow{\pa}_\b} \psi
= \phi \psi - \phi \overleftarrow{\pa}_\a C^{\a \b} \overrightarrow{\pa}_\b
\psi - \frac12 C^2 \pa^2\phi {\pa}^2 \psi
\label{star}
\eeq
where we have defined $C^2 =  C^{\a \b}C_{\a \b}$.
Since the covariant derivatives (\ref{DQ}) are still derivations for this
product, if we define (anti)chiral superfields as usual the classes of
(anti)chirals are still closed. However, this product explicitly breaks
the $\bar{Q}$--supersymmetry, being defined in terms of noncovariant spinor
derivatives.

Before closing this section we note that, if we were to use the antichiral 
representation for charges and covariant derivatives 
(basically by interchanging the definitions of $Q$'s and $D$'s 
in (\ref{DQ})) we would still obtain
a NAC generalization of superspace described by the algebra (\ref{NAC1}) 
but in this case the algebra of derivatives would get deformed as in
(\ref{NC30}), while  the susy charges would satisfy the ordinary 
anticommutation rules. Moreover, since in this representation $D_{\a} = 
\partial_{\a}$, the $\ast$--product would be naturally defined in terms
of covariant derivatives, so avoiding explicit breaking of supersymmetry.
One might conclude that in this representation supersymmetry is not
broken at all. 
However, the modification of the anticommutation relations between
covariant derivatives makes it difficult to proceed and consistently define
(anti)chiral representations. 
This issue certainly requires more investigation.

\section{The $N=1/2$ WZ model}

Given the $N=(1/2,0)$ {\em euclidean} superspace and the algebra of the smooth 
functions defined on it endowed with the $\ast$--product (\ref{star}),
we may define NAC supersymmetric field theories by promoting ordinary 
lagrangians to NAC ones where the ordinary products have been substituted with 
$\ast$--products. The simplest model we can consider is the NAC generalization
of the WZ model described by the action
\bea
&& S = \int d^8z \Phib \Phi - \frac{m}{2} \int d^6z \Phi^2
- \frac{\bar{m}}{2} \int d^6\bar{z} \Phib^2
\nonumber \\
&& ~~~- \frac{g}{3} \int d^6z \Phi \ast \Phi \ast \Phi - \frac{\bar{g}}{3} \int
d^6\bar{z} \Phib \ast \Phib \ast \Phib 
\label{action} 
\eea
This action is generically complex since no h.c. relations are assumed for 
fields, masses and couplings.

Performing the expansion of the star product as in (\ref{star}) and 
neglecting total superspace derivatives, the cubic interaction terms 
reduce to the usual WZ interactions augmented by the nonsupersymmetric 
component term $\frac{g}{6}C^2 \int d^4 x F^3 $ \cite{Seiberg}, 
where $F = D^2 \Phi|$. 
The action can be written as
\bea 
&& S =
\int d^8z \Phib \Phi - \frac{m}{2} \int d^6z \Phi^2 - \frac{\bar{m}}{2} \int
d^6\bar{z} \Phib^2 - \frac{g}{3} \int d^6z \Phi^3  - \frac{\bar{g}}{3} \int
d^6\bar{z} \Phib^3
\nonumber \\
&& ~~~~+ \frac{g}{6} \int d^8z U (D^2 \Phi)^3 
\label{action2} 
\eea 
where we have introduced the external, constant spurion superfield 
\cite{spurion} $U = \theta^2 \thb^2 C^2$ in order to deal with a
well--defined superspace expression for the extra term proportional to the
NAC parameter. This allows us to use standard supergraphs techniques to perform
perturbative calculations. 

The action in
components reads ($\Phi| = \phi$, $D_\a \Phi| = \psi_\a$, $D^2 \Phi| = F$ and
analogously for the antichiral components) 
\bea &&S ~=~ \int d^4x \Big[ \phi
\Box \bar{\phi} + F \bar{F}
-GF -\bar{G}\bar{F} + \frac{g}{6} C^2 F^3  \nonumber\\
&&~~~~~~~~~~~~~~~ + \psi^\a i \pa_\a^{\adot} \bar{\psi}_{\adot} - \frac{m}{2}
\psi^\a \psi_\a - \frac{\bar{m}}{2} \bar{\psi}^{\adot} \bar{\psi}_{\adot}
- g \phi \psi^\a \psi_\a - \bar{g} \phib \bar{\psi}^{\adot} \bar{\psi}_{\adot}
\Big]
\label{components}
\eea
where we have defined
\beq 
G = m \phi + g \phi^2 \qquad \quad
\bar{G}= \bar{m} \phib + \bar{g} \phib^2 
\label{GGbar}
\eeq 
The auxiliary fields $F$ and
$\bar{F}$ satisfy the algebraic equations of motion (EOM) 
\beq F = \bar{G}
\qquad , \qquad \bar{F} = G - \frac{g}{2} C^2 F^2 = G- \frac{g}{2} C^2
\bar{G}^2 
\label{EOM} 
\eeq

In \cite{us2} the perturbative evaluation of the effective action up to 
two loops has been performed. Here we report the basic results and refer
the reader to that reference for details of the calculation.

The procedure we have applied is the following:
We have performed quantum--background splitting by setting 
$\Phi \rightarrow \Phi +
\Phi_q$ and integrating out the quantum fluctuations $\Phi_q$. From the 
expansion of the action (\ref{action2}) we have read the Feynman rules
for vertices and propagators. It is important to note that the propagators
are the ordinary ones since the quadratic part of the action is not modified
by the $\ast$--product. Instead, two new quadratic and cubic vertices 
appear from the $U$ term. 
At a given loop order we have drawn all supergraph configurations
with the corresponding chiral and antichiral derivatives from the
vertices and the propagators. Then we have performed $D$--algebra to reduce
the supergraphs to ordinary momentum diagrams. 
We have worked in dimensional regularization and minimal subtraction
scheme. We have used BPHZ renormalization techniques which amounts to  
start with the classical action written in terms of renormalized 
quantities and order by order perform
the subtraction of subdivergences directly on the diagrams.
Finally, in the counterterms we need add to the action to remove divergent 
contributions, we have made repeated use of the EOM (\ref{EOM}) for the 
auxiliary field $F$. At any loop order this is justified by the important 
observation that, due to the particular form of the propagators, the 
insertion of counterterms proportional to $\bar{G}= 
\bar{m} \phib + \bar{g} \phib^2$ that one performs into higher loop diagrams
in order to cancel subdivergences is {\em equivalent} to the insertion of 
counterterms proportional to $F$ (see \cite{us2} for details). 

Applying this strategy, the main results we have obtained are the 
following: At one loop we have the ordinary self--energy $\bar{\Phi}\Phi$ 
divergent diagram which induces a wave function renormalization, plus 
two new divergent contributions proportional to $F^2$ and $F^3$
coming from diagrams with one $U$ insertion and one external $\Phi$,
and external $\Phi$, $\bar{\Phi}$ and $\Phi$, $\bar{\Phi}$, $\bar{\Phi}$, 
respectively.
The $F^3$ divergence induces a renormalization of the spurion $U$, whereas
the $F^2$ does not have a classical counterpart and makes the model 
(\ref{action}) not renormalizable.    
Thus we have considered a modified action with the addition of $F^2$ 
and $F$ terms (once we have $F^2$ it is obvious that we will generate
tadpoles) which in superspace language reads
\beq
S_{r}= S + k_1 \bar{m}^4 \int d^8 z U D^2 \Phi + k_2 \bar{m}^2 \int d^8z
U (D^2 \Phi)^2 
\label{WZaction}
\eeq   
with $S$ given in (\ref{action2}).
Starting from this action, we have computed the divergent 
contributions up to two loops and found the following results:
\begin{itemize}
\item Divergent diagrams contain at most one $U$ vertex
\item Divergences are always logarithmic
\item The antiholomorphic part of the action does not get renormalized. 
Moreover, the ordinary part of the action (the one independent of 
$C^{\alpha \beta}$) does not receive contributions proportional to the 
anticommutation parameter
\item In components the general structures of the divergent terms are
\end{itemize} 
\begin{equation}
C^2 \int d^4x \Big[ a_0 F + a_1 F \bar{G} + a_2 
F^2 \bar{G} + a_3 F \bar{G}^2  + a_4 F + a_5 F^2 + a_6 F^3 \Big]
\end{equation} 

where $\bar{G}$ is given in (\ref{GGbar}). 

Now, using 
the classical equation of motion, $F = \bar{G}$, all the 
divergent terms assume the form $F, F^2, F^3$. 
In conclusion, we have proved that the counterterms $F, F^2, F^3$ 
(or their superspace expressions 
in terms of $U$ superfield) are sufficient to renormalize the theory 
(\ref{WZaction}) up to two loops.

In \cite{us2} the two--loop beta functions for the couplings of the theory
have been also computed. Even if we expect them to be affected by scheme 
dependence
it is interesting to note that nontrivial fixed points for the NAC 
parameter might exist.

\section{Conclusions} 
 
In these proceedings we have reviewed the results of \cite{us2} 
concerning the two--loop renormalizability of the NAC WZ model described 
by the action (\ref{WZaction}). The main result is the appearance of the 
extra $F$ and $F^2$ terms in the classical action which, as shown in 
\cite{us2}, are sufficient to make the theory renormalizable at two loops.
Further investigations have been carried on in \cite{BFR2} where it has
been proven that the addition of these extra terms is sufficient to 
make the theory renormalizable at any loop order. A basic ingredient of 
the proof is the existence of two global U(1) (pseudo)symmetries which
constrain the structure of the counterterms.
 
We note that the $F^2$ term cannot be written as the $\ast$--product of 
anything. Therefore,
one might worry about the presence of this extra term as deforming the
definition of $\ast$--product at quantum level. 
A possible interpretation of this term
has been discussed in 
\cite{BFR3}. In a string theory context it would be nice to 
understand the origin of these extra contributions.
  
Our approach could be suitable for performing perturbative calculations 
in NAC Yang--Mills theories (indications about renormalizability of those 
theories are contained in \cite{rey2,rey}). 
In that case, an explicit expression
for the NAC action has been worked out in components in the WZ gauge
and one--loop calculations have been done \cite{SYM}. In order to implement
ordinary superspace techniques and push the calculations beyond one loop
it should be necessary to find an expression for the NAC action in the gauge
of superspace rather than in the WZ gauge.  

Other interesting generalizations of our analysis would concern susy 
field theories defined in superspaces where also the bosonic coordinates 
would be noncommuting. 

\vskip 30pt

{\bf Acknowledgments} This work has been supported in part by INFN, MURST
and the European Commission RTN program HPRN--CT--2000--00131 in which S.P. 
and A.R. are associated to the University of Padova. The work of M.T.G. is 
supported by NSF Grant No. PHY-0070475 and by NSERC Grant No. 204540.

\end{document}